\documentclass[11pt]{article}
\usepackage{graphicx}

\def\sintwob    {\ensuremath{\sin 2 \beta}}

\def\fish    {\ensuremath{\it F}}

\def\btodstarrho {\ensuremath{\B \to {D}^{\star}\:\rho}}

\def\btojpsipi {\ensuremath{\B^0 \to {J}/\psi\:\piz}}
\def\btojpsiks {\ensuremath{\B^0 \to {J}/\psi\:\ks}}

\def\ks  {\ensuremath{{K}_{S}}}

\def\jpsi {\ensuremath{J}/\psi}
\def\jpsitoll {\ensuremath{\jpsi\rightarrow\lplus\lminus}}
\def\jpsitoee {\ensuremath{\jpsi\rightarrow\eplus\eminus}}
\def\jpsitomm {\ensuremath{\jpsi\rightarrow\muplus\muminus}}
\def\lplus {\ensuremath{\ell^+}}
\def\lminus {\ensuremath{\ell^-}}
\def\eplus {\ensuremath{e^+}}
\def\eminus {\ensuremath{e^-}}
\def\muplus {\ensuremath{\mu^+}}
\def\muminus {\ensuremath{\mu^-}}

\def\CPV {\ensuremath{CPV}}
\def\A {\ensuremath{A}}
\def\C {\ensuremath{C}}
\def\S {\ensuremath{S}}
\def\mes   {\ensuremath{m_{ES}}}
\def\de   {\ensuremath{\Delta E}}
\def\dt   {\ensuremath{\Delta t}}

\def\dz   {\ensuremath{\Delta z}}

\def\d    {\ensuremath{d}}
\def\u    {\ensuremath{u}}
\def\s    {\ensuremath{s}}

\def\bb   {\ensuremath{B\Bbar}}

\def\piz {\ensuremath{ \pi^0 }}

\def\de      {\ensuremath{\Delta E}}
\def\cerenkov{$\check{\rm C}{\rm erenkov}$ }

\newcommand{\BABARPubYear}    {05}

\newcommand{\BABARConfNumber} {16}
\newcommand{\SLACPubNumber} {11362}

\RequirePackage{xspace}

\usepackage{relsize}
\def\babar{\mbox{\slshape B\kern-0.1em{\smaller A}\kern-0.1em
    B\kern-0.1em{\smaller A\kern-0.2em R}}}

\def\epem       {\ensuremath{e^+e^-}\xspace}

\def\q     {\ensuremath{q}\xspace}

\def\qqbar {\ensuremath{q\overline q}\xspace}
\def\u     {\ensuremath{u}\xspace}

\def\d     {\ensuremath{d}\xspace}

\def\s     {\ensuremath{s}\xspace}

\def\c     {\ensuremath{c}\xspace}

\def\piz   {\ensuremath{\pi^0}\xspace}

\def\Kbar  {\kern 0.2em\overline{\kern -0.2em K}{}\xspace}

\def\Kz    {\ensuremath{K^0}\xspace}
\def\Kzb   {\ensuremath{\Kbar^0}\xspace}
\def\KzKzb {\ensuremath{\Kz \kern -0.16em \Kzb}\xspace}
\def\Kp    {\ensuremath{K^+}\xspace}
\def\Km    {\ensuremath{K^-}\xspace}

\def\KpKm  {\ensuremath{\Kp \kern -0.16em \Km}\xspace}

\def\Dbar    {\kern 0.2em\overline{\kern -0.2em D}{}\xspace}

\def\Dz      {\ensuremath{D^0}\xspace}
\def\Dzb     {\ensuremath{\Dbar^0}\xspace}
\def\DzDzb   {\ensuremath{\Dz {\kern -0.16em \Dzb}}\xspace}
\def\Dp      {\ensuremath{D^+}\xspace}
\def\Dm      {\ensuremath{D^-}\xspace}

\def\DpDm    {\ensuremath{\Dp {\kern -0.16em \Dm}}\xspace}

\def\B       {\ensuremath{B}\xspace}
\def\Bbar    {\kern 0.18em\overline{\kern -0.18em B}{}\xspace}

\def\BB      {\ensuremath{B\Bbar}\xspace} 
\def\Bz      {\ensuremath{B^0}\xspace}
\def\Bzb     {\ensuremath{\Bbar^0}\xspace}
\def\BzBzb   {\ensuremath{\Bz {\kern -0.16em \Bzb}}\xspace}
\def\Bu      {\ensuremath{B^+}\xspace}
\def\Bub     {\ensuremath{B^-}\xspace}

\def\BpBm    {\ensuremath{\Bu {\kern -0.16em \Bub}}\xspace}

\def\BorBbar    {\kern 0.18em\optbar{\kern -0.18em B}{}\xspace}
\def\DorDbar    {\kern 0.18em\optbar{\kern -0.18em D}{}\xspace}
\def\KorKbar    {\kern 0.18em\optbar{\kern -0.18em K}{}\xspace}

\def\jpsi     {\ensuremath{{J\mskip -3mu/\mskip -2mu\psi\mskip 2mu}}\xspace}

\mathchardef\Upsilon="7107
\def\Y#1S{\ensuremath{\Upsilon{(#1S)}}\xspace}

\mathchardef\Deltares="7101
\mathchardef\Xi="7104
\mathchardef\Lambda="7103
\mathchardef\Sigma="7106
\mathchardef\Omega="710A

\def\Deltabar{\kern 0.25em\overline{\kern -0.25em \Deltares}{}\xspace}
\def\Lbar{\kern 0.2em\overline{\kern -0.2em\Lambda\kern 0.05em}\kern-0.05em{}\xspace}
\def\Sigbar{\kern 0.2em\overline{\kern -0.2em \Sigma}{}\xspace}
\def\Xibar{\kern 0.2em\overline{\kern -0.2em \Xi}{}\xspace}
\def\Obar{\kern 0.2em\overline{\kern -0.2em \Omega}{}\xspace}
\def\Nbar{\kern 0.2em\overline{\kern -0.2em N}{}\xspace}
\def\Xb{\kern 0.2em\overline{\kern -0.2em X}{}\xspace}

\def\mes        {\mbox{$m_{\rm ES}$}\xspace}

\newcommand{\tev}{\ensuremath{\mathrm{\,Te\kern -0.1em V}}\xspace}
\newcommand{\gev}{\ensuremath{\mathrm{\,Ge\kern -0.1em V}}\xspace}
\newcommand{\mev}{\ensuremath{\mathrm{\,Me\kern -0.1em V}}\xspace}
\newcommand{\kev}{\ensuremath{\mathrm{\,ke\kern -0.1em V}}\xspace}
\newcommand{\ev}{\ensuremath{\mathrm{\,e\kern -0.1em V}}\xspace}
\newcommand{\gevc}{\ensuremath{{\mathrm{\,Ge\kern -0.1em V\!/}c}}\xspace}
\newcommand{\mevc}{\ensuremath{{\mathrm{\,Me\kern -0.1em V\!/}c}}\xspace}
\newcommand{\gevcc}{\ensuremath{{\mathrm{\,Ge\kern -0.1em V\!/}c^2}}\xspace}
\newcommand{\mevcc}{\ensuremath{{\mathrm{\,Me\kern -0.1em V\!/}c^2}}\xspace}

\def\mus  {\ensuremath{\rm \,\mus}\xspace}

\def\ps   {\ensuremath{\rm \,ps}\xspace}

\def\mus        {\ensuremath{\,\mu{\rm s}}\xspace}    
\def\ps         {\ensuremath{{\rm \,ps}}\xspace}  

\def\to                 {\ensuremath{\rightarrow}\xspace}

\newcommand{\stat}{\ensuremath{\mathrm{(stat)}}\xspace}
\newcommand{\syst}{\ensuremath{\mathrm{(syst)}}\xspace}

\def\pep2{PEP-II}

\def\gsim{{~\raise.15em\hbox{$>$}\kern-.85em
          \lower.35em\hbox{$\sim$}~}\xspace}
\def\lsim{{~\raise.15em\hbox{$<$}\kern-.85em
          \lower.35em\hbox{$\sim$}~}\xspace}

\def\CP                {\ensuremath{C\!P}\xspace}
\def\C       {\ensuremath{C}\xspace}

\def\deltat{\ensuremath{{\rm \Delta}t}\xspace}
\def\deltamd{\ensuremath{{\rm \Delta}m_d}\xspace}

\xspace

\newcommand{\jprlBase}       {Phys.\ Rev.\ Lett.\xspace}
\newcommand{\jprBase}        {Phys.\ Rev.\xspace}
\newcommand{\jplBase}        {Phys.\ Lett.\xspace}
\newcommand{\nimBaseA}       {Nucl.\ Instr.\ Methods Phys.\ Res., Sect.\ A\xspace}

\newcommand{\npBase}         {Nucl.\ Phys.\xspace}

\newcommand{\nima}      [1]  {\nimBaseA~{\bf #1}}

\newcommand{\np}        [1]  {\npBase\ {\bf #1}}
\newcommand{\plb}       [1]  {\jplBase\ B~{\bf #1}}

\newcommand{\jprl}      [1]  {\jprlBase\ {\bf #1}}
\newcommand{\pr}        [1]  {\jprBase\ {\bf #1}}
\newcommand{\jprd}      [1]  {\jprBase\ D~{\bf #1}}

\def\jetset74   {\mbox{\tt Jetset \hspace{-0.5em}7.\hspace{-0.2em}4}\xspace}

\long\def\inst#1{\par\nobreak\kern 4pt\nobreak
    {\it #1}\par\vskip 10pt plus 3pt minus 3pt}

\begin{document}
{\pagestyle{empty}


\begin{flushright}
\babar-CONF-\BABARPubYear/\BABARConfNumber \\
SLAC-PUB-\SLACPubNumber \\
July 2005 \\
\end{flushright}

\par\vskip 5cm

\begin{center}
\Large \bf Measurements of the Branching Fraction and Time-Dependent \CP\-Asymmetries of $\btojpsipi$ Decays
\end{center}
\bigskip

\begin{center}
\large The \babar\ Collaboration\\
\mbox{ }\\
\today
\end{center}
\bigskip \bigskip

\begin{center}
\large \bf Abstract
\end{center}
We present measurements of the branching fraction and time-dependent \CP\ asymmetries 
in\newline $\btojpsipi$ decays based on (231.8 $\pm$ 2.6) $\times$ 10$^{6}$ 
$\Upsilon$(4S)$\rightarrow\bb$ decays collected with the \babar\ detector at the PEP-II 
asymmetric-energy \B\ factory at SLAC during the years 1999-2004. We obtain a branching fraction 
$\cal{B}$($\btojpsipi$) = (1.94 $\pm$ 0.22 \stat $\pm$ 0.17 \syst)$\times$ 10$^{-5}$. We also
measure the \CP\ asymmetry parameters \C\ = $-$0.21 $\pm$ 0.26 \stat $\pm$ 0.09 \syst 
and \S\ = $-$0.68 $\pm$ 0.30 \stat $\pm$ 0.04 \syst.
All results presented in this paper are preliminary. 
\vfill
\begin{center}
Presented at the 
International Europhysics Conference On High-Energy Physics (HEP 2005),
7/21---7/27/2005, Lisbon, Portugal
\end{center}

\vspace{1.0cm}
\begin{center}
{\em Stanford Linear Accelerator Center, Stanford University, 
Stanford, CA 94309} \\ \vspace{0.1cm}\hrule\vspace{0.1cm}
Work supported in part by Department of Energy contract DE-AC03-76SF00515.
\end{center}

\newpage
} 

\begin{center}
\small

The \babar\ Collaboration,
\bigskip

B.~Aubert,
R.~Barate,
D.~Boutigny,
F.~Couderc,
Y.~Karyotakis,
J.~P.~Lees,
V.~Poireau,
V.~Tisserand,
A.~Zghiche
\inst{Laboratoire de Physique des Particules, F-74941 Annecy-le-Vieux, France }
E.~Grauges
\inst{IFAE, Universitat Autonoma de Barcelona, E-08193 Bellaterra, Barcelona, Spain }
% ...

\end{center}\newpage

\section{Introduction}
\label{sec:Introduction}
Charge conjugation-parity (\CP) violation in the $B$ meson system has been
established by the \babar~\cite{babar-stwob-prl}
and Belle~\cite{belle-stwob-prl} collaborations.
The Standard Model (SM) of electroweak interactions describes \CP\ violation
as a consequence of a complex phase in the
three-generation Cabibbo-Kobayashi-Maskawa (CKM) quark-mixing
matrix~\cite{ref:CKM}. Measurements of \CP\ asymmetries in
the proper-time distribution of neutral $B$ decays to
\CP\ eigenstates containing a charmonium and $K^{0}$ meson provide
a precise measurement of $\sintwob$~\cite{BCP}, where
$\beta$ is $\arg \left[\, -V_{\rm cd}^{}V_{\rm cb}^* / V_{\rm td}^{}V_{\rm tb}^*\, \right]$ and
the $V_{ij}$ are CKM matrix elements. 

\par
\begin{figure}[h!]
\begin{center}
\includegraphics[height=6cm]{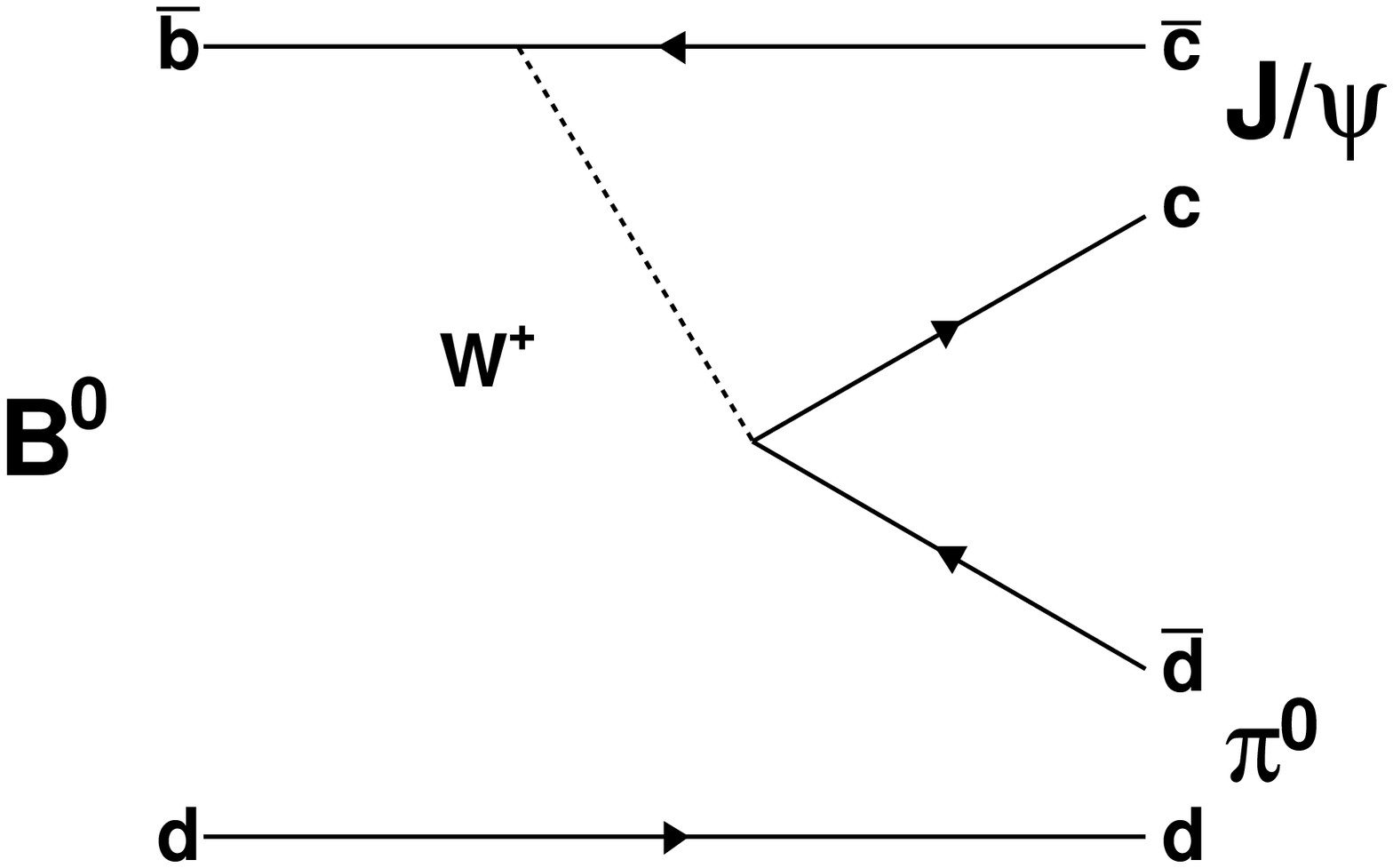}
\includegraphics[height=6cm]{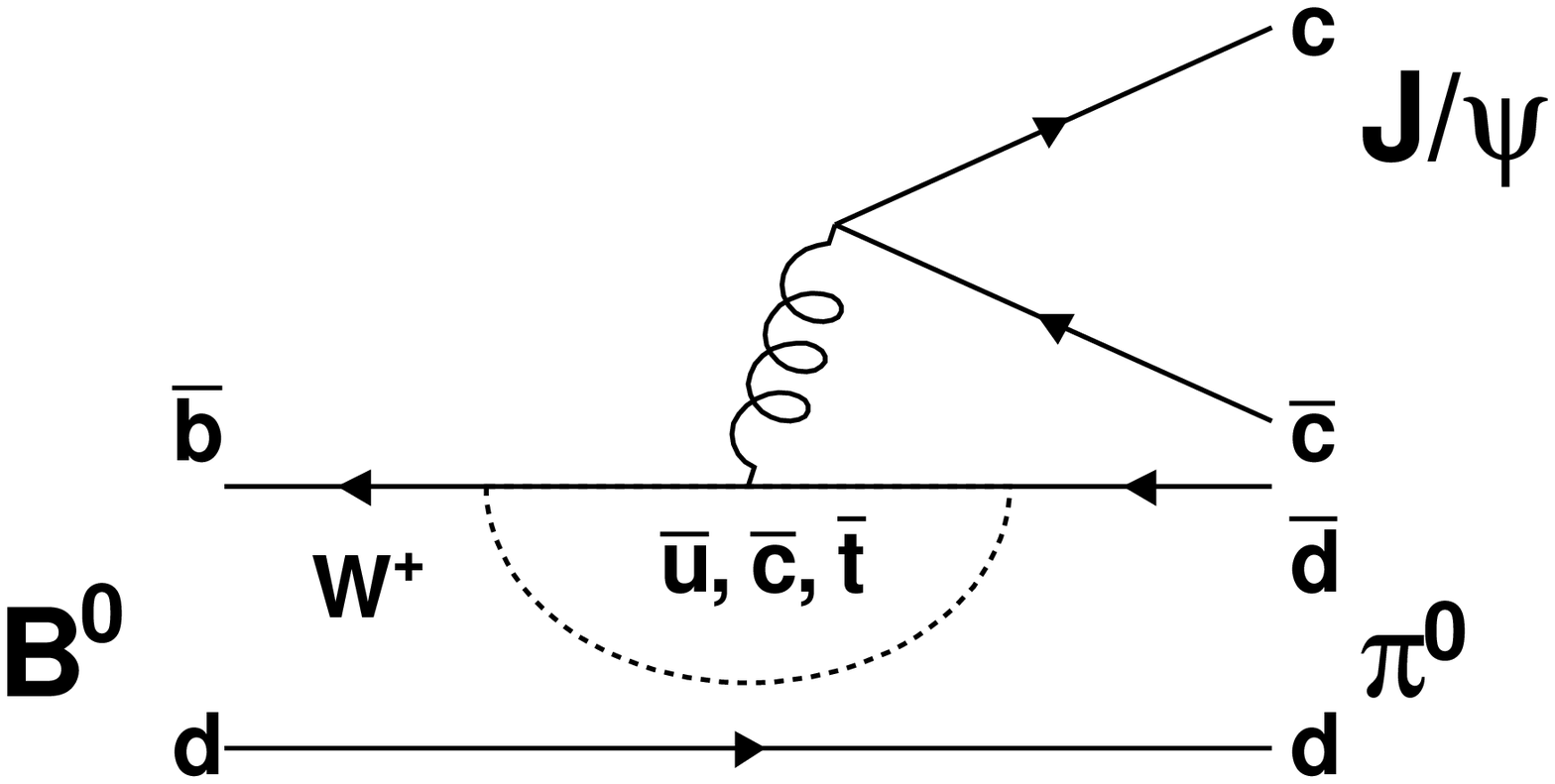}
\caption{Feynman diagrams of the color suppressed tree (left) and gluonic penguin (right) amplitudes contributing to the \btojpsipi\ decay.}
\label{fig:feynman}
\end{center}
\end{figure}

 The decay $\btojpsipi$ is a \CP-even Cabibbo-suppressed 
${b \rightarrow c\mskip 2mu \overline c \mskip 2mu d}$ transition for which, in the absence of 
loop (penguin) amplitudes, the SM predicts that the two \CP asymmetry coefficients, $\S$, 
the interference between mixing and decay, and $\C$, the direct \CP asymmetry,
 are $\S$ = $-\sintwob$, and $\C = 0$. $\S$ and $\C$ are defined as:
\begin{eqnarray}
S \equiv \frac{{2\mathop{\cal I\mkern -2.0mu\mit m}}
  \lambda}{1+|\lambda|^2}
\mskip 50mu {\rm and} \mskip 50mu
C \equiv \frac{1  - |\lambda|^2 }{1+|\lambda|^2},
\label{equation-sandc}
\end{eqnarray}
where $\lambda$ is a complex parameter that depends on both the \Bz-\Bzb oscillation 
amplitude and the amplitudes describing \Bz and \Bzb decays to the $\jpsi\piz$ final state.
The tree and penguin amplitudes expected to dominate 
this decay are shown in Figure~\ref{fig:feynman}. \newline

The ${b \rightarrow c\mskip 2mu \overline c \mskip 2mu d}$
tree amplitude has the same weak phase as the 
${b \rightarrow c\mskip 2mu \overline c \mskip 2mu s}$ modes
(e.g. the \CP-odd decay ${B^0 \rightarrow J\mskip -3mu/\mskip -2mu\psi\mskip 2mu
K^0_{\scriptscriptstyle S}}$). The ${b \rightarrow c\mskip 2mu \overline c \mskip 2mu d}$ 
penguin amplitudes have a different weak phase than the tree amplitude.
If there is a significant penguin amplitude in $\btojpsipi$, then one will 
measure a value of \S\ that differs from $-\sintwob$, and a value of $\C$ 
that differs from zero~\cite{grossman}. 

In this paper, we present an update of previous \babar\ measurements
of the branching fraction and time-dependent \CP\-violating
(\CPV) asymmetries in $\btojpsipi$~\cite{ref:babarbf,ref:babarcp}. The preliminary results presented here are
 obtained using 210.6 fb$^{-1}$ of data. \babar\ and Belle have both previously presented 
measurements of the $\btojpsipi$ branching fraction
using $\Upsilon$(4S)$\rightarrow\bb$ decays. These are: 
\newpage
\begin{table}[h]
\begin{center}
\begin{tabular}{cccc}
\babar: & (2.0 $\pm$ 0.6 \stat $\pm$ 0.2 \syst)$\times$10$^{-5}$ & (from 20.7 fb$^{-1}$) &~\cite{ref:babarbf},\\
 Belle:  & (2.3 $\pm$ 0.5 \stat $\pm$ 0.2 \syst)$\times$10$^{-5}$ & (from 29.4 fb$^{-1}$) &~\cite{ref:bellebf}.\\
\end{tabular}
\end{center}
\end{table}
Both the \babar\ and Belle notations in denoting the magnitude of the direct \CP\ asymmetry, 
where \C (\babar) = $-$\A\ (Belle), the previous measurements of \S\ and \C\ ($\A$) are:
\begin{table}[h]
\begin{center}
\begin{tabular}{ccccc}
\babar: &  \S\ = 0.05 $\pm$ 0.49 $\pm$ 0.16  &  \C\ = 0.38 $\pm$ 0.41 $\pm$ 0.09   & (from 81.1 fb$^{-1}$) ~\cite{ref:babarcp},\\ 
 Belle:  &  \S\ = $-$0.72 $\pm$ 0.42 $\pm$ 0.09 & \A\ = $-$0.01 $\pm$ 0.29 $\pm$ 0.03 & (from 140.0 fb$^{-1}$)~\cite{ref:bellecp}.\\
\end{tabular}
\end{center}
\end{table}
\section{The \babar\ detector and dataset}
\label{sec:babar}
The data used in this analysis were collected with the \babar\ detector
at the \pep2\ asymmetric e$^{+}$e$^{-}$storage ring from 1999 to 2004. This 
represents a total integrated luminosity of 210.6 fb$^{-1}$ taken at the 
$\Upsilon$(4S) resonance (onpeak), corresponding to a sample of 231.8 $\pm$ 2.6 
million $\BB$ pairs. An additional 21.6 fb$^{-1}$ of data, collected 
at approximately 40 MeV below the $\Upsilon$(4S) resonance, is used to study
background from \epem\to\qqbar ($\q = \u,\d,\s,\c$) continuum events.

The \babar\ detector is described elsewhere~\cite{ref:babar}. Surrounding the
interaction point is a 5 layer double-sided silicon vertex tracker (SVT) which
provides precise reconstruction of track angles and \B\ decay vertices. A
40 layer drift chamber (DCH) surrounds the SVT and provides measurements
of the transverse momenta for charged particles. Both the SVT and the DCH
 operate in a 1.5 T solenoidal magnetic field. Charged hadron 
identification is achieved through measurements of particle energy-loss 
($dE/dx$) in the tracking system and \cerenkov angle obtained 
from a detector of internally reflected \cerenkov light (DIRC).  This is surrounded by a 
segmented CsI(Tl) electromagnetic calorimeter (EMC) which provides photon 
detection, electron identification, and is used to reconstruct neutral 
hadrons.  Finally, the instrumented flux return (IFR) of the magnet allows 
discrimination of muons from pions.

\section{Analysis Method}
\label{sec:Analysis}
We study $\btojpsipi$ decays in $\bb$ candidate events from combinations
of $\jpsitoll$ ($\ell$ = e, $\mu$) and $\piz\rightarrow\gamma\gamma$ candidates.
A detailed description of the $\jpsitoll$ selection can be found elsewhere~\cite{ref:babarbf}. 
For the $\jpsitoee$ ($\jpsitomm$) channel, the invariant mass of the lepton pair is 
required to be between $3.06$ and $3.12 \gevcc$ ($3.07$ and $3.13 \gevcc$). 

We form $\piz\to\gamma\gamma$ candidates with an invariant mass
$100 < m_{\gamma\gamma} < 160$ {\mevcc} from pairs of photon
candidates which have been identified as clusters in the EMC. These
clusters are required to be isolated from any charged tracks, carry
a minimum energy of 30{\mev}, and have a lateral energy distribution 
 consistent with that of a photon. Each $\piz$ candidate is required to
have a minimum energy of 200{\mev} and is constrained
to the nominal mass~\cite{ref:pdg2004}.

The $\btojpsipi$ candidates ($B_{rec}$) are reconstructed from these $\jpsi$ and 
$\piz$ candidates and constrained to originate from the
\epem{} interaction point using a geometric fit. 
Finally, the \jpsi\ and \piz\ candidates are combined using 4-momentum addition.


We use two kinematic variables, $\mes$ and $\de$, in order to isolate the signal.\newline
$\mes=\sqrt{(\sqrt{s}/2)^2-(p_B^{\rm *})^2}$ is the beam-energy substituted mass,
where $\sqrt{s}$ is the center-of-mass (CM) energy, and therefore $\sqrt{s}/2$ is the beam
energy in the CM frame. $p_B^{\rm *}$ is the \B-candidate momentum in 
the CM frame. \de\ is the difference between the \B-candidate energy and the
beam energy in the \epem\ CM frame. We require $\mes > 5.2 \gevcc$ and $|\de| < 0.3 \gev$.


\par A significant source of background is due to \epem\to\qqbar ($\q = \u,\d,\s,\c$) 
continuum events.  We combine several kinematic and topological variables 
into a Fisher discriminant (\fish) to provide additional separation between signal
 and continuum. The three variables $L_0$, $L_2$ and $\cos$($\theta_{H}$) are inputs to \fish.
$L_0$ and $L_2$ are the zeroth- and second-order Legendre polynomial moments;
$L_0 = \sum_i |{\bf p}^{\rm *}_i|$ and $L_2 = \sum_i |{\bf p}^{\rm *}_i| \hspace{0.5mm} \frac{3 \cos^2\theta_i - 1}{2}$,
where ${\bf p}^{\rm *}_i$ are the CM momenta for the tracks and neutral
calorimeter clusters that are not associated with the signal candidate. The
$\theta_i$ are the angles between ${\bf p}^{\rm *}_i$ and the thrust axis of
the signal candidate. $\theta_{H}$ is the angle between one of the leptons 
and the $\B$ candidate in the $\jpsi$ rest frame.

\par We use multivariate algorithms to identify signatures of \B\ decays that determine (tag)
the flavor of the decay of the other \B in the event ($\B_{tag}$) to be either a \Bz or \Bzb. The flavor 
tagging algorithm used is described in more detail elsewhere~\cite{ref:babar2004}.
In brief, we define seven mutually exclusive tagging categories.  These are 
(in order of decreasing signal purity) {\tt Lepton}, 
{\tt Kaon 1}, {\tt Kaon 2}, {\tt Kaon-Pion}, {\tt Pion}, {\tt Other}, 
and {\tt No-Tag}.
The total effective tagging efficiency of this algorithm is ($30.5 \pm 0.4$)\%.

\par The decay rate $f_+$ ($f_-$) of \Bz decays to a \CP eigenstate, 
when $B_{tag}$ is a  \Bz (\Bzb), is:
\begin{equation}
f_{\pm}(\dt) = \frac{e^{-\left|\dt\right|/\tau_{\Bz}}}{4\tau_{\Bz}} [1
\pm S\sin(\deltamd\deltat) \mp \C\cos(\deltamd\dt)],  
\label{equation-ff}
\end{equation}
where \dt\ is the difference between the proper decay times of 
the $B_{rec}$ and $B_{tag}$ mesons, $\tau_{\Bz}$ = 1.536 $\pm$ 0.014 ps is the \Bz\ lifetime
and \deltamd\ = (0.502 $\pm$ 0.007) $\times$10$^{-12}$s is the \Bz-\Bzb\ oscillation frequency~\cite{ref:pdg2004}. 
The decay width difference between the \Bz\ mass eigenstates is assumed to be zero. 

\par The time interval \dt\ is calculated from the measured separation \dz\ between
the decay vertices of $B_{rec}$ and $B_{tag}$ along the collision axis ($z$).
The vertex of $B_{rec}$ is from the lepton tracks that come from the $J/\psi$
and the vertex of $B_{tag}$ is constructed from the remaining tracks in the event that do
not belong to $B_{rec}$, whilst using constraints from the beam spot location
and the $B_{rec}$ momentum.  We accept events with $|\dt|<20 \ps$ whose
uncertainty is less than $2.5 \ps$.  

\par After all of the selection cuts mentioned above have been applied, the average 
multiplicity is approximately 1.1, indicating some events still have multiple candidates. 
In these events, we randomly choose one candidate to be used in the fit. 
After this step, the signal efficiency is $22.0 \%$ and a total of 1318 onpeak events are selected.

\par In addition to signal and continuum background events, there are also \B\ backgrounds present 
in the data after applying the selection cuts above.  We divide the \B\ backgrounds into the
following types: (i) \btojpsiks, (ii) generic neutral \B\ meson decays, and (iii) generic charged
\B\ meson decays. From Monte Carlo (MC) we expect $153\pm 9$, $68 \pm 14$ and $314 \pm 63$ events of these 
background types, respectively. The generic  neutral \B\ meson decays do not include
signal or \btojpsiks\ events. The generic \B\ decay backgrounds are dominated by contributions from
$\B \to J/\psi X$ (inclusive charmonium final states). In particular the generic charged
\B\ meson decay backgrounds are dominated by $\B^{\pm} \to J/\psi \rho^{+}$ decays. The \btojpsiks\ background was 
studied separately since it is a well understood decay with respect to time-dependent analysis.

\par We perform an extended unbinned maximum likelihood fit to the $\B$ candidate sample, where the 
discriminating variables used in the fit are \mes, \de, \fish\ and \dt.  The values of the signal yield, 
\S\ and \C are simultaneously extracted.

The signal \mes\ distribution is described by a Gaussian with an 
exponential tail~\cite{ref:crystalball}.  We parameterise the \mes\ distribution for continuum 
and neutral generic \B\ background with a phase space distribution~\cite{ref:argus}.
As there are significant correlations between \mes\ and \de\ for the 
charged generic \B\ background, we parameterise these variables with
two-dimensional non-parametric probability density functions (PDFs). We use two-dimensional 
non-parametric PDFs when describing the \mes-\de\ distribution for \btojpsiks.
The \de\ distribution for signal events is modeled with a Gaussian with an
exponential tail on the negative side to model energy leakage in the EMC, plus a polynomial contribution. The \de\ distribution for 
continuum and neutral generic \B\ background are described by polynomials.
The \fish\ distributions for the signal and the backgrounds are described by 
a Gaussian with different widths above and below the mean (a bifurcated Gaussian).

The signal decay rate distribution of Equation~\ref{equation-ff} is modified to account for dilution coming from 
incorrectly assigning the flavor of $B_{tag}$ and is convoluted with a triple 
Gaussian resolution function, whose core width is about 1.1 \ps.
The decay rate distribution for \B\ backgrounds is similar to that for signal.
To account for their mis-reconstruction, the generic \B\ backgrounds are assigned 
an effective lifetime instead of their respective measured \B\ lifetimes.
When evaluating systematic uncertainties, we allow for \CP\ violation 
in the generic \B background. This is described later in the text. 
The decay rate distribution for \btojpsiks\ is the same as that for signal 
and accounts for the known level of \CP\ violation in that decay.  The continuum 
background is modeled with a prompt lifetime component convoluted with a triple Gaussian 
resolution function.

The results from the fit are 109 $\pm$ 12 \stat signal events, with \S\ = $-$0.68 $\pm$ 0.30 \stat
and \C = $-$0.21 $\pm$ 0.26 \stat. We also obtain for the aforementioned
mutually exclusive tagging categories, the following numbers of
continuum events: N$_{\tt{Lepton}}$ = 17 $\pm$ 5,
N$_{\tt{Kaon1}}$ = 38 $\pm$ 8, N$_{\tt{Kaon2}}$ = 101 $\pm$ 12, 
N$_{\tt{KaonPion}}$ = 102 $\pm$ 12, N$_{\tt{Pion}}$ = 115 $\pm$ 12, 
N$_{\tt{Other}}$ = 94 $\pm$ 11 and N$_{\tt{NoTag}}$ = 227 $\pm$ 17.

\begin{figure}[h!]
\begin{center}
\vspace{2cm}
\includegraphics[height=5.54cm]{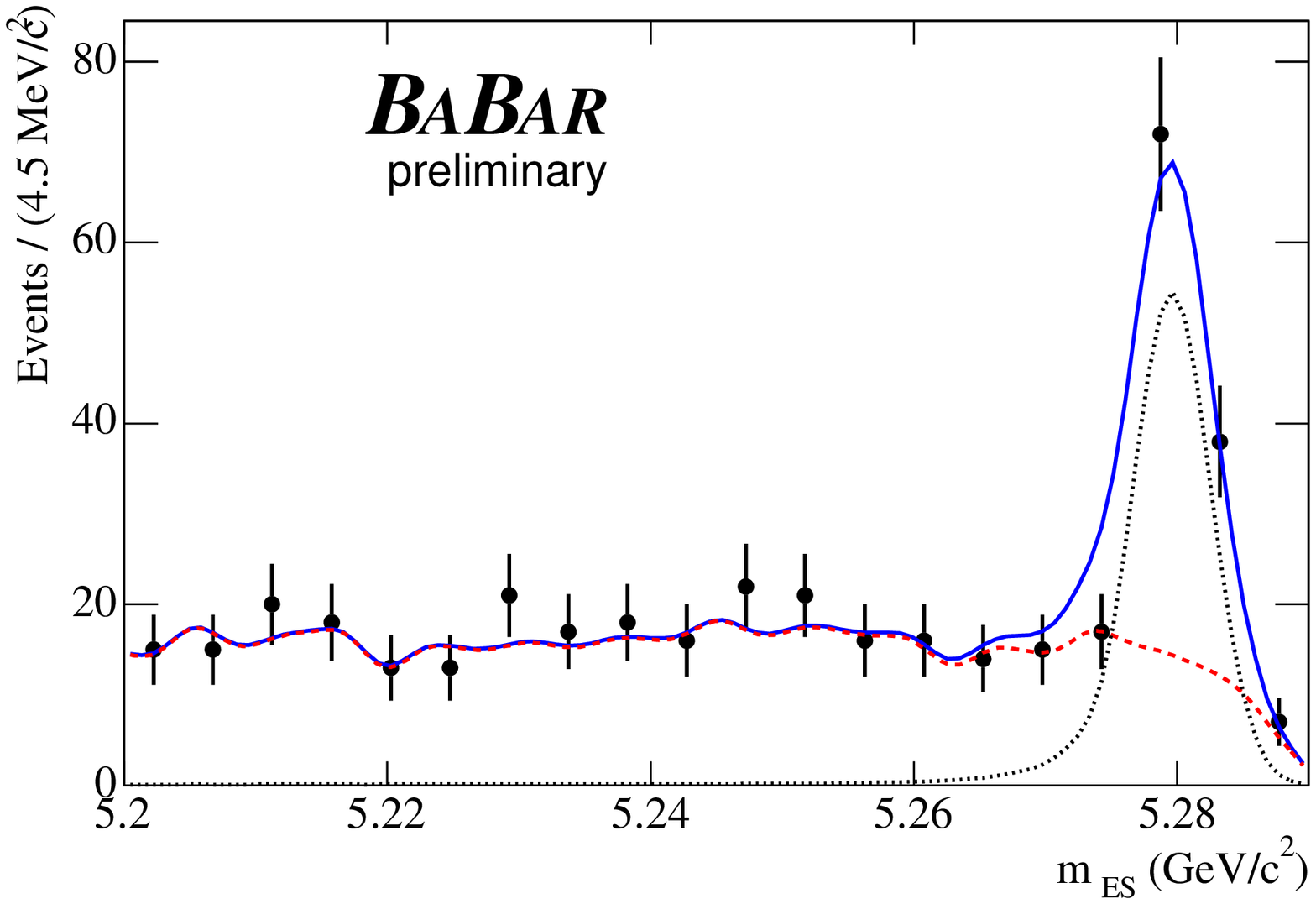}
\includegraphics[height=5.54cm]{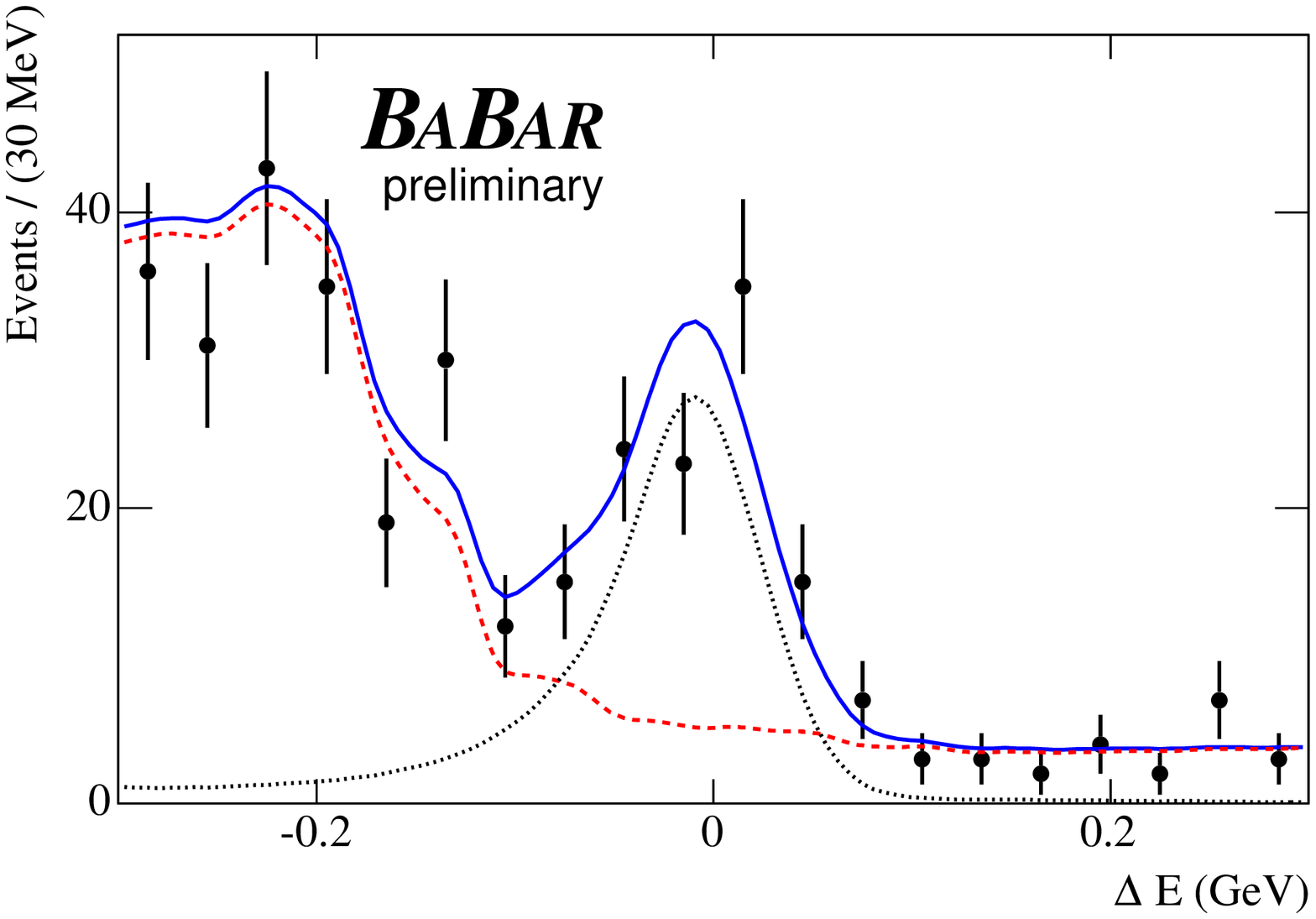}
\includegraphics[height=5.54cm]{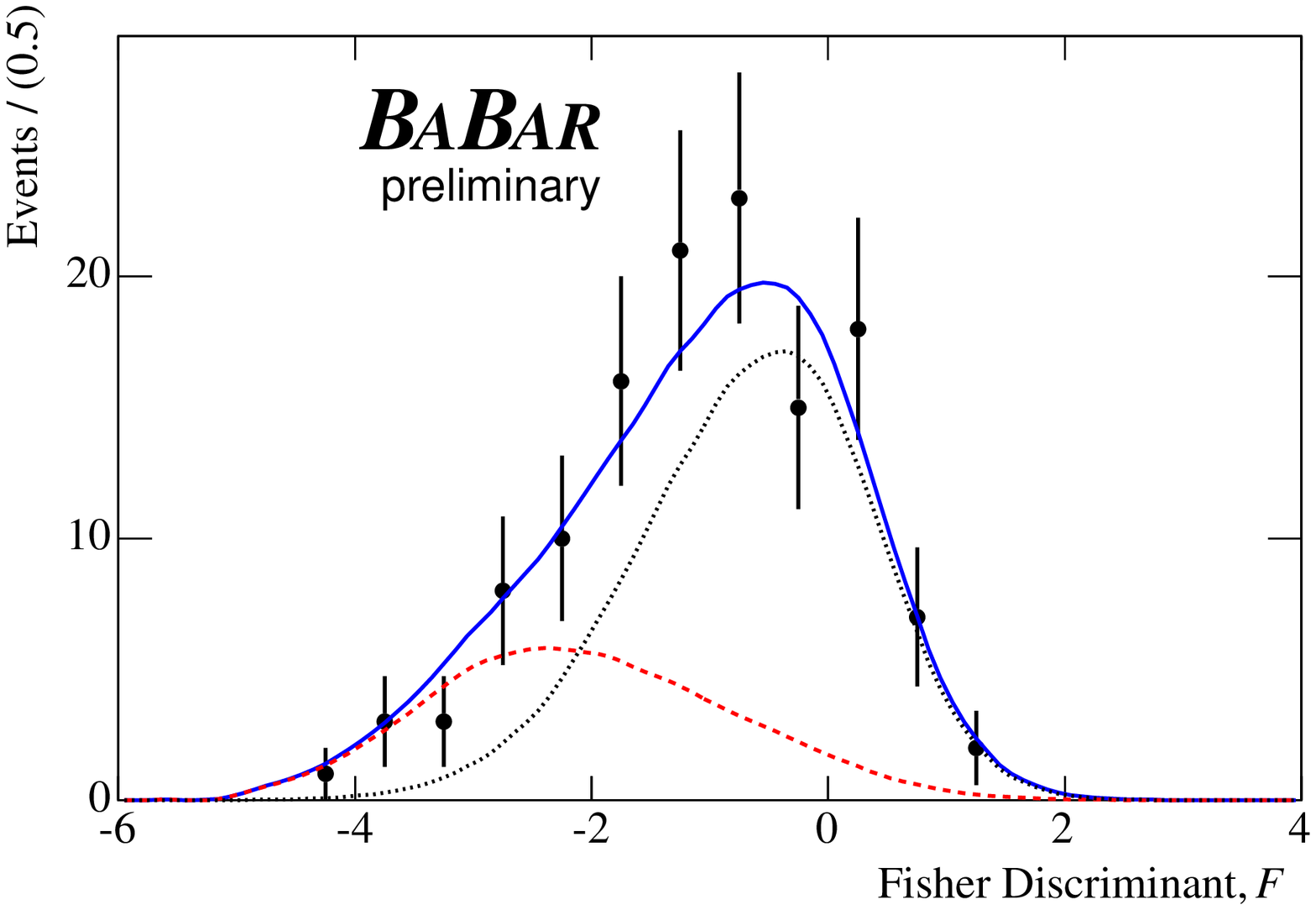}
\caption{Signal enhanced distributions of \mes\ (top), \de\ (center) and
 \fish\ (bottom) for the data (black points). The (blue) solid line represents the 
total likelihood, the (red) dashed line is the sum of the backgrounds and the (black)
 dotted line is the signal. The undulations in the background model are the 
result of limited MC statistics available for defining the two-dimensional 
non-parametric PDFs.}
\label{fig:projection}
\end{center}
\end{figure}
\begin{figure}[h!]
\begin{center}
\vspace{2cm}
\includegraphics[height=8.5cm]{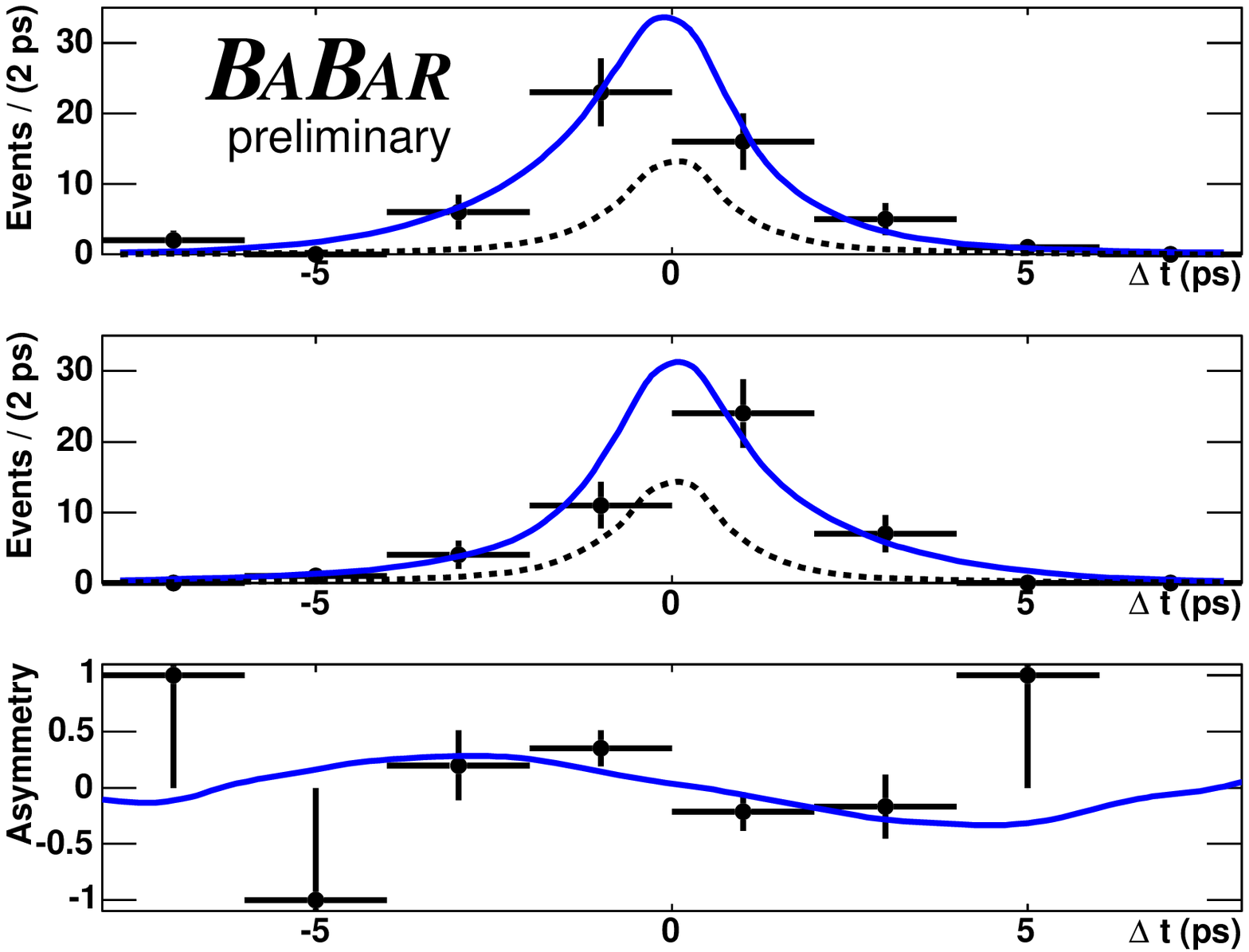}
\caption{The \deltat\ distribution for a sample of events enriched in signal for $\Bz$ (top) and $\Bzb$ (middle) tagged events. 
The dotted lines are the sum of backgrounds and the solid lines are the sum of signal and backgrounds.  
The time-dependent \CP asymmetry (see text) is also shown (bottom), where the curve is the measured asymmetry.}
\label{fig:asym}
\end{center}
\end{figure}

Figure~\ref{fig:projection} shows the distributions of \mes, \de, and
\fish\ for the data.  In these plots the signal has been enhanced by 
cutting on $|\de| < 0.1 \gev$ for the \mes plot, 
$\mes > 5.275\gevcc$ for the \de\ plot and by applying
 both of these constraints for the \fish\ plot.  
After applying these requirements to the signal (background) samples that
 are used in the fit, they are reduced to a relative size of 83.1\% (24.3\%), 
85.0\% (21.1\%) and 73.1\% (2.8\%) for the \mes, \de, and \fish\ distributions respectively.

Figure~\ref{fig:asym} shows the \deltat\ distribution for \Bz and \Bzb
tagged events. The time-dependent decay-rate asymmetry 
$[  N (\deltat) - \overline{N}(\deltat) ] / [  N (\deltat) + \overline{N}(\deltat) ]$
is also shown, where $N$ $(\overline{N})$ is the decay-rate for \Bz(\Bzb) tagged events
and the decay-rate takes the form of Equation~\ref{equation-ff}.

\section{Systematic Studies}
\label{sec:Systematics}
Table~\ref{table:systematicerrors} summarises the systematic uncertainties on 
the signal yield, \S\ and \C. Each entry in the table indicates one systematic
effect and these are added in quadrature to give the totals presented. These 
include the uncertainty due to the PDF parameterisation, 
evaluated by fixing both the signal and the background PDF parameters to their nominal values
and varying them within uncertainties;
the effect of SVT mis-alignment; the uncertainty due to knowledge of the Lorentz boost and
z-scale of the tracking system, and knowledge of the event-by-event beam spot position. 

\begin{table}[h]
\caption{Contributions to the systematic errors on the signal yield, \S\ and \C.  Additional systematic 
uncertainties that are applied only to the branching fraction are discussed in the text.~\label{table:systematicerrors}}
\begin{center}
\begin{tabular}{|c|c|c|c|}\hline
Contribution                    & Signal yield       & \S\                  & \C\                 \\\hline\hline
                                &                    &                      &                     \\
SVT mis-alignment               & $-$                & $\pm${0.002}         & $\pm${0.002}        \\
Boost and z-scale               & $^{+0.08}_{-0.16}$ & $\pm${0.004}         & $\pm${0.001}        \\
Beam spot position              & $-$                & $\pm${0.007}         & $\pm${0.002}        \\
PDF parameterisation            & $^{+3.21}_{-2.88}$ & $\pm${0.013}         & $^{+0.009}_{-0.012}$\\
Fit bias                        & $\pm$3.00          & $\pm${0.030}          & $\pm${0.060}         \\
Generic $\B$ background yields  & $\pm$3.52          & $\pm${0.003}         & $\pm${0.020}         \\ 
Choice of 1D v 2D PDFs          & $\pm$2.92          & $\pm${0.020}          & $\pm${0.002}        \\
\CP\ content of \B\ background  & $^{+0.40}_{-0.26}$ & $^{+0.020}_{-0.018}$ & $\pm${0.058}        \\
\CP\ background lifetime        & $\pm$0.67          & $\pm${0.010}          & $\pm${0.010}         \\
Tagging efficiency asymmetry    & $\pm$0.02          & $\pm${0.000}          & $\pm${0.020}         \\
Tag-side interference           & $-$                & $\pm${0.004}         & $\pm${0.014}        \\
Fisher data/MC comparison       & $\pm${0.70}        & $\pm${0.004}         & $\pm${0.004}        \\\hline      
                                &                    &                      &                     \\
Total                           & $^{+6.43}_{-6.26}$ & $^{+0.044}_{-0.043}$ & $\pm${0.093}\\
                                &                    &                      &                     \\\hline
\end{tabular}
\end{center}
\end{table}

\par The uncertainty coming from the fit bias is estimated by performing ensembles of mock 
experiments using signal MC which is generated using the GEANT based \babar\ 
MC simulation~\cite{ref:geant}, embedded into MC samples of background generated from the 
PDF.  The deviation from input values is added in quadrature to the error on the 
deviation in order to obtain the fit bias uncertainty. Most, but not all of the inclusive 
charmonium final states which dominate the generic \B\ background, are precisely known 
from previous measurements. Their yields are then fixed in the fit. 
As a crosscheck, we allow the backgrounds to vary in the fit to data to validate the expected yields 
and to provide a systematic uncertainty. We also apply an additional systematic uncertainty 
to account for neglecting the small correlation between $\mes$ and $\de$ in signal and 
neutral generic \B\ background events.

\par In order to evaluate the uncertainty coming from  \CP\ violation in the \B\ background, we
have allowed \S\ and \C\ to vary between $+1$ and $-1$ for the neutral generic \B\ background,
and for the direct \CP asymmetry to vary between $+0.5$ and $-0.5$ for the charged 
generic \B\ background. The \CP\ parameters in \btojpsiks\ are varied within current 
experimental knowledge~\cite{ref:babar2004}. 

\par The generic \B\ background uses an effective lifetime in the nominal fit. We replace 
this with the \B\ lifetime to evaluate the systematic error due to \CP\ background lifetime.
There is also a small asymmetry in the tagging efficiency between $\Bz$ and $\Bzb$ tagged events, for
which a systematic uncertainty is evaluated. We also study the possible interference between the 
suppressed $\bar b\to \bar u c \bar d$ amplitude with the favored $b\to c \bar u d$ amplitude 
for some tag-side $B$ decays~\cite{ref:dcsd}. The difference in the distribution of \fish\
between data and MC is evaluated with a large sample of $\btodstarrho$ decays.

There are additional systematic uncertainties that contribute only to the branching fraction.  These come from
uncertainties in charged particle identification (5.2\%), \piz\ meson reconstruction (3\%), the $\jpsitoll$
branching fractions (2.4\%), tracking efficiency (1.2\%) and the number of \B\ meson
pairs (1.1\%).  The systematic error contribution from MC statistics is negligible.  

\section{Summary}
\label{sec:Summary}
The 109 $\pm$ 12 signal events correspond to a preliminary branching fraction of
\vspace{-0.8cm}
\begin{center}
$$
{\cal{B}}(\btojpsipi) = (1.94 \pm 0.22 \stat \pm 0.17 \syst)\times 10^{-5}
$$
\end{center}\vspace{-0.2cm}
which is consistent with previous measurements from the \B\ Factories.
We determine the preliminary \CP asymmetry parameters
\vspace{-0.9cm}
\begin{center}
\begin{eqnarray}
\C\ = -0.21 \pm 0.26 \stat \pm 0.09 \syst, \nonumber \\
\S\ = -0.68 \pm 0.30 \stat \pm 0.04 \syst, \nonumber
\end{eqnarray}
\end{center}\vspace{-0.2cm}
where the correlation between \S\ and \C\ is 8.3\%. The value of \S\ is consistent with
SM expectations for a tree-dominated ${b \rightarrow c\mskip 2mu \overline c \mskip 2mu d}$ 
transition of \S\ = $-\sintwob$ and \C\ = 0.


\section{Acknowledgments}
\label{sec:Acknowledgments}
We are grateful for the 
extraordinary contributions of our \pep2\ colleagues in
achieving the excellent luminosity and machine conditions
that have made this work possible.
The success of this project also relies critically on the 
expertise and dedication of the computing organizations that 
support \babar.
The collaborating institutions wish to thank 
SLAC for its support and the kind hospitality extended to them. 
This work is supported by the
US Department of Energy
and National Science Foundation, the
Natural Sciences and Engineering Research Council (Canada),
Institute of High Energy Physics (China), the
Commissariat \`a l'Energie Atomique and
Institut National de Physique Nucl\'eaire et de Physique des Particules
(France), the
Bundesministerium f\"ur Bildung und Forschung and
Deutsche Forschungsgemeinschaft
(Germany), the
Istituto Nazionale di Fisica Nucleare (Italy),
the Foundation for Fundamental Research on Matter (The Netherlands),
the Research Council of Norway, the
Ministry of Science and Technology of the Russian Federation, and the
Particle Physics and Astronomy Research Council (United Kingdom). 
Individuals have received support from 
CONACyT (Mexico),
the A. P. Sloan Foundation, 
the Research Corporation,
and the Alexander von Humboldt Foundation.



\begin{thebibliography}{99}

\bibitem{babar-stwob-prl}
\babar\ Collaboration, B.\ Aubert {\em et al.},
\jprl{89}, 201802 (2002).

\bibitem{belle-stwob-prl}
BELLE Collaboration, K.\ Abe {\em et al.},
\jprd{66}, 071102 (2002).

\bibitem{ref:CKM}
N.~Cabibbo, Phys.~Rev.~Lett.~{\bf 10}, 531 (1963);\\
M.~Kobayashi and T.~Maskawa, Prog.\ Th.\ Phys.\ {\bf 49}, 652 (1973).

\bibitem{BCP}
A.B.~Carter and A.I.~Sanda, \pr {\bf D23}, 1567 (1981);\\
I.I.~Bigi   and A.I.~Sanda, \np {\bf B193}, 85 (1981).

\bibitem{grossman}
Y.~Grossman and M.~Worah,
\plb{395}, 241 (1997).

\bibitem{ref:babarbf}
\babar\ Collaboration, B.\ Aubert {\em et al.},
\jprd{65}, 032001 (2002).

\bibitem{ref:babarcp}
\babar\ Collaboration, B.\ Aubert {\em et al.},
\jprd{91}, 061802 (2003).

\bibitem{ref:bellebf}
BELLE Collaboration, K.\ Abe {\em et al.},
\jprd{67}, 032003 (2002).


\bibitem{ref:bellecp}
BELLE Collaboration, K.\ Abe {\em et al.},
\jprl{93}, 261801 (2004).

\bibitem{ref:babar}
\babar\ Collaboration, B.\ Aubert {\em et al.},
\nima{479}, 1 (2002).

\bibitem{ref:pdg2004}
Particle Data Group, 
S.~Eidelman {\em et al.},
\plb{592}, 1 (2004).

\bibitem{ref:babar2004}
\babar\ Collaboration, B.\ Aubert {\em et al.},
\jprl{94}, 161803 (2005).

\bibitem{ref:crystalball}
Crystal Ball Collaboration, D.~Antreasyan {\em et al.}, Crystal Ball Note 321 (1983).

\bibitem{ref:argus}
ARGUS Collaboration, H.\ Albrecht {\em et al.},
\plb{241}, 278 (1990).

\bibitem{ref:geant}
Geant4 Collaboration, S. Agostinelli {\em et al.},
\nima{506}, 250 (2003).

\bibitem{ref:dcsd}
O.~Long, M.~Baak, R.~N.~Cahn, D.~Kirkby, 
\jprd{68}, 034010 (2003).

\end{thebibliography}
\end{document}